\documentclass[aps,amsmath,amssymb,twocolumn,floatfix]{revtex4-2}

\usepackage{graphicx}
\usepackage{dcolumn}
\usepackage{bm}

\begin{document}

\title{The 2D metallic triangular lattice antiferromagnet CeCd$_{3}$P$_3$}

\author{Jeonghun Lee$^{1}$, Anja Rabus$^{1}$, N. R. Lee-Hone$^{1}$, D. M. Broun$^{1,2}$,  and Eundeok Mun$^{1}$}

\affiliation{$^{1}$Department of Physics, Simon Fraser University, Burnaby, BC V5A1S6, Canada}
\affiliation{$^{2}$Canadian Institute for Advanced Research, Toronto, ON, MG5 1Z8, Canada}

\date{\today}

\begin{abstract}

Single crystals of $R$Cd$_{3}$P$_{3}$ ($R$ = La and Ce) have been investigated by magnetization, electrical resistivity, Hall coefficient, and specific heat. Magnetization measurements of CeCd$_{3}$P$_{3}$ demonstrate clear quasi-2D magnetic behavior. Electrical resistivity and Hall coefficient measurements suggest that $R$Cd$_{3}$P$_{3}$ compounds are low carrier density metallic systems, in strong contrast to an earlier study of polycrystalline material. Specific heat and electrical resistivity measurements of CeCd$_{3}$P$_{3}$ reveal a high temperature (structural) phase transition at $T_{s} = 127$~K and antiferromagnetic ordering below $T_{N} = 0.41$~K. Upon applying magnetic field in the easy-plane ($H {\parallel} ab$) the magnetic ordering temperature increases to 0.43~K at $H \sim 15$~kOe, demonstrating partial lifting of the magnetic frustration. The large electronic specific heat persists in an unusually wide range of temperature above  $T_{N}$, due to the frustrated spins. The observation of conventional metallic behavior in the electrical resistivity suggests that the $f$-electrons in CeCd$_{3}$P$_{3}$ undergo negligible hybridization with the conduction electrons. Thus, CeCd$_{3}$P$_{3}$ may be a model system for exploring the complex interplay between magnetic frustration and RKKY physics on a low carrier density Ce triangular lattice.

\end{abstract}


\maketitle


\section{Introduction}

The ground states of geometrically frustrated insulating magnets exhibit a range of unconventional order parameters.\cite{Ramirez1994, Lacroix2011, Starykh2015} In low dimensional quantum magnets, competing magnetic exchange interactions give rise to strong frustration accompanied by enhanced quantum fluctuations. For such systems, frustration may prevent the magnet from forming long range order, leading to magnetically liquid states.\cite{Balents2010, Lee2008, Savary2017, Zhou2017} These ``spin liquids'' come in different forms, depending on the type of magnetic exchange interaction ($e.g.$, Heisenberg, Dzyaloshinskii-Moriya, or Kitaev) and the lattice geometry ($e.g.$, square, triangular, Kagome, honeycomb, or pyrochlore).\cite{Lacroix2011, Balents2010, Savary2017} In particular, spins on two-dimensional (2D) triangular lattices, interacting antiferromagnetically via XY or Heisenberg exchange, provide an excellent opportunity to study various ground states, and have strong potential for realizing the spin-liquid state in 2D.\cite{Wannier1950, Mermin1966, Anderson1973, Teitel1983, Miyashita1984, Kawamura1984, Lee1984, Yosefin1985, Lee1986} Until now, most such spin systems have been insulating.  Finding examples in which the spin-liquid state coexists with itinerant conduction electrons remains challenging, but offers the possibility of revealing highly novel electronic states.

For triangular lattice (TL) magnets with 4$f$-electrons,  spin-orbit entanglement strongly enhances quantum fluctuations and promotes a liquid ground state characterized by highly anisotropic interactions between moments.\cite{Knolle2016, Hu2015, Li2016, Iqbal2016, Gong2017, Zhu2018, Rau2018} In the absence of spin-orbit coupling, it has been shown that the frustration is partially lifted by forming a planar 120$^{\textrm{o}}$ spin structure with strong magnetic anisotropy.\cite{Kawamura1985, Chubukov1991, Seabra2011, Yamamoto2014, Schmidt2017} Examples of $f$-electron materials with 2D TL structures include spin-gapped YbAl$_{3}$C$_3$\cite{Ochiai2007, Kato2008, Ochiai2010, Hara2012}, spin-liquid systems YbMgGaO$_{4}$ \cite{Li2015, Li2015A, Shen2016, Li2016A, Paddison2017} and NaYbS$_{2}$\cite{Baenitz2018, Sichelschmidt2019}, and easy-plane antiferromagnets CeCd$_{3}$P$_3$ and CeCd$_{3}$As$_3$.\cite{Higuchi2016, Liu2016} Recently, a putative quantum spin-liquid state in which magnetic order remains absent and magnetic excitations persist down to low temperatures has been claimed for 4$f$-electron insulating TL magnets such as YbMgGaO$_{4}$\cite{Shen2016} and NaYbS$_{2}$,\cite{Baenitz2018} where an effective $J_\mathrm{eff} = 1/2$ spin moment can be realized due to strong spin-orbital coupling in conjunction with the crystalline electric field (CEF) effect. For metallic materials, containing Ce and Yb elements,  the physical properties are associated with the competition between Kondo hybridization and RKKY interactions.\cite{Doniach1977, Stewart1984} However, rich behavior can also be driven by magnetic frustration, which promotes complex ordering and might even lead to a quantum spin liquid state under some circumstances.\cite{Coleman2010, Si2010, Coleman2010A, Custers2012, Kim2013, Mun2013, Tokiwa2013, Fritsch2014} Clearly, then, it is desirable to uncover new $f$-electron metals satisfying the conditions for magnetic frustration.

In this report, we present physical properties of single crystals of $R$Cd$_{3}$P$_{3}$ ($R = $ La and Ce). At room temperature, $R$Cd$_{3}$P$_{3}$ materials adopt the hexagonal ScAl$_{3}$C$_{3}$-type structure (space group $P6_{3}/\mathrm{mmc}$), in which the Ce triangular layers are well separated by the Cd and P atoms and form a 2D, geometrically frustrated TL in the $ab$-plane, with the Ce$^{3+}$ atoms having trigonal point symmetry.\cite{Higuchi2016, Nientiedt1999, Stoyko2011, Banda2018} The results of magnetization, electrical and Hall resistivity, and specific heat measurements of single crystal CeCd$_{3}$P$_{3}$ indicate strongly anisotropic quasi-2D magnetism associated with low carrier density metallic behavior; an emergent spin-orbit-entangled doublet ground state of Ce at low temperatures; a high temperature (structural) phase transition at $T_{s} = 127$~K; and low temperature antiferromagnetic ordering at $T_{N} = 0.41$~K. Previously, polycrystalline CeCd$_{3}$P$_{3}$  was reported to be a semiconductor with a band gap of $\sim$0.75~eV, with measurements of magnetic susceptibility revealing no magnetic ordering down to 0.48~K.\cite{Higuchi2016} Similarly, the isostructural system CeZn$_{3}$P$_{3}$ was reported as showing semiconducting behavior with a relatively small band gap.\cite{Yamada2010, Kitagawa2013, Kitagawa2016}

\section{Experiments}

\begin{figure}
\centering
\includegraphics[width=1\linewidth]{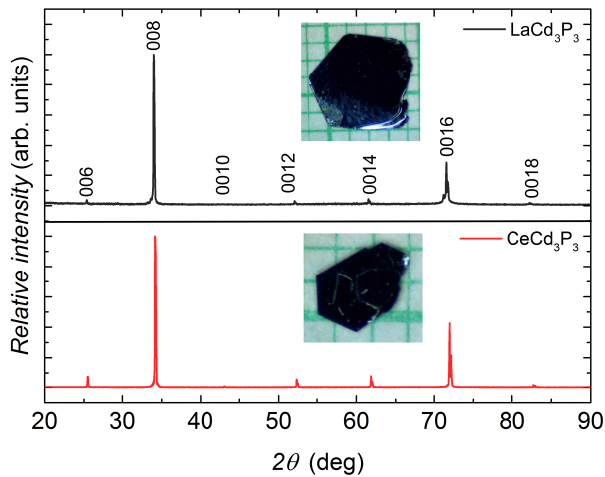}
\caption{(a) (b) Single crystal X-ray patterns for $R$Cd$_{3}$P$_{3}$ ($R$ = La and Ce). Insets show photographs of LaCd$_{3}$P$_{3}$ and CeCd$_{3}$P$_{3}$ single crystals on a 1~mm grid scale. (c) Crystal structure of $R$Cd$_{3}$P$_{3}$. (d) Local coordination environments of $R$-site.}
\label{FIG1}%
\end{figure}

Single crystals of $R$Cd$_{3}$P$_{3}$ ($R$ = La and Ce) were prepared by high temperature ternary melt.\cite{Canfield1992} The as-grown single crystals have hexagonal morphology and form very thin $ab$-plane platelets, reflecting their layered structure, as shown in Fig.~\ref{FIG1}. The samples have been characterized  using powder X-ray diffraction (XRD) in a Rigaku MiniFlex instrument at room temperature. The XRD pattern contains no indications of impurity phases. Analysis of the powder XRD patterns shows that samples crystallize in the hexagonal ScAl$_{3}$C$_{3}$-type structure \mbox{($P$6$_{3}$/mmc, 194)} with lattice parameters \mbox{$a$ = 4.2767~\AA}~and $c$ = 20.9665~\AA~for CeCd$_{3}$P$_{3}$ and \mbox{$a$ = 4.2925~\AA}~and $c$ = 21.0763~\AA~for LaCd$_{3}$P$_{3}$,  consistent with earlier work.\cite{Higuchi2016}  As seen in the platelet XRD results in Fig.~\ref{FIG1}, only (0, 0, $\ell$) reflection peaks are detected, indicating that the crystallographic $c$-axis is perpendicular to the planes.

Magnetization was measured as a function of temperature, from 1.8 to 300~K, and magnetic field, up to 70~kOe, using a Quantum Design (QD) Magnetic Property Measurement System (MPMS). Four-probe ac resistivity measurements were performed in a QD Physical Property Measurement System (PPMS). Hall resistivity measurements were performed in a four-wire geometry, for which the magnetic field directions were reversed to remove magnetoresistance effects due to voltage-contact misalignment. Specific heat was measured by the relaxation method down to $T$~=~0.37~K in a QD PPMS. For the dc transport measurements, samples were prepared by attaching Pt wires using silver paste. Due to the high contact resistance, of the order of 50~$\Omega$ at room temperature, we were not able to measure dc resistivity at low temperatures. Thus, microwave surface resistance measurements were performed below 5~K, at a frequency of 202~MHz.

\section{Results}

\begin{figure}
\centering
\includegraphics[width=1\linewidth]{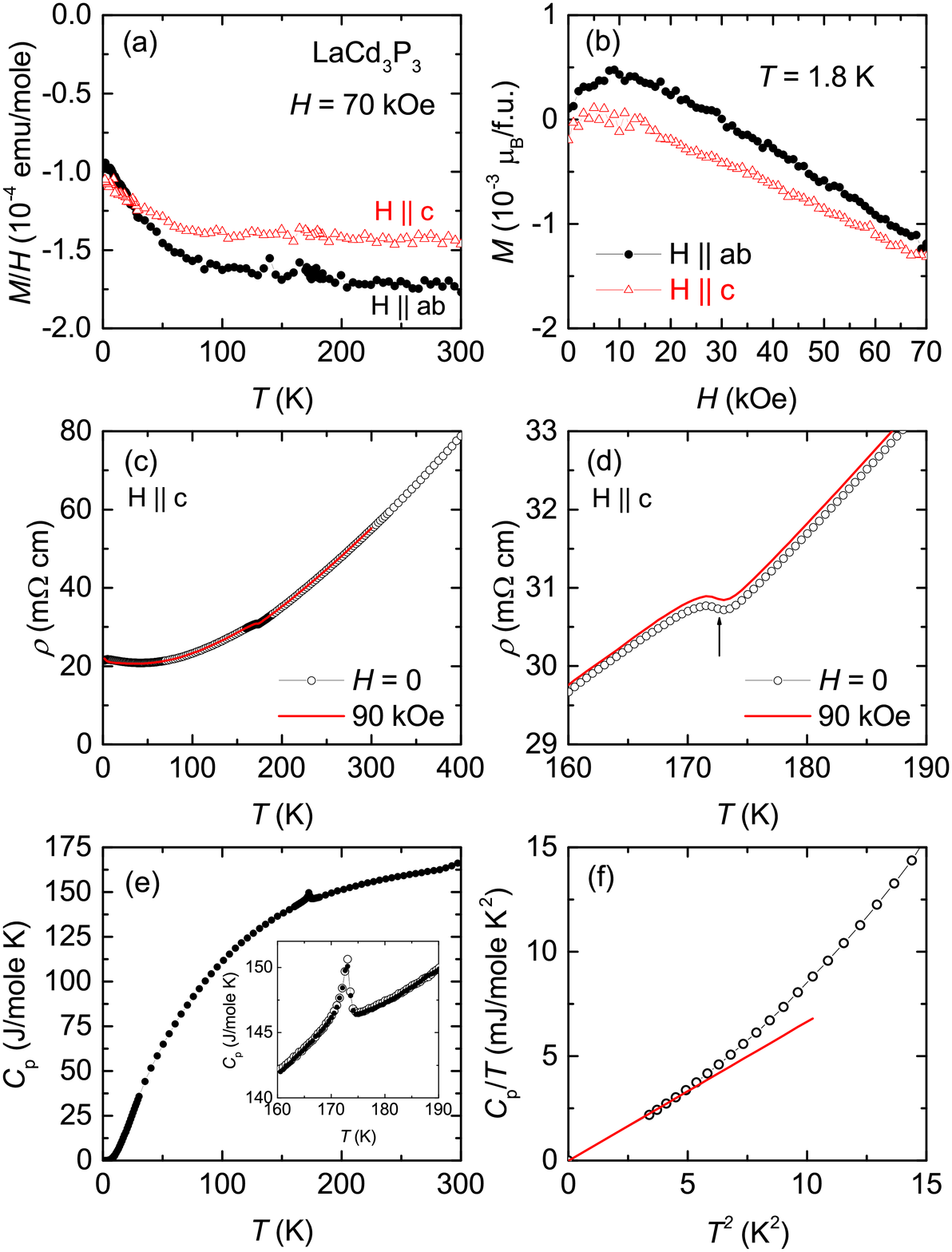}
\caption{Physical properties of LaCd$_{3}$P$_{3}$. (a) Magnetic susceptibility $M/H$ at $H = 70$~kOe for $H {\parallel} ab$ and $H {\parallel} c$. (b) Magnetization isotherm $M(H)$ at $T = 1.8$~K. (c) Electrical resistivity $\rho(T)$ at $H = 0$ and 90~kOe. (d) The data from (c) for 160~K $< T <$ 190~K. Vertical arrow indicates a minimum in $d\rho(T)/dT$. (e) Specific heat $C_{p}$. The inset shows an enlarged plot near the phase transition. Open and closed symbols are the data taken on warming and cooling, respectively. (f) $C_{p}/T$ vs.\ $T^2$. The solid line represents the linear extrapolation of $C_{p}/T$ below 2.1~K.}
\label{FIG2}%
\end{figure}

Figure~\ref{FIG2}(a) shows the temperature dependence of magnetic susceptibility, $\chi(T) = M/H$, of LaCd$_{3}$P$_{3}$. $\chi(T)$ displays temperature independent, diamagnetic behavior down to  roughly 100~K. As temperature decreases $\chi(T)$ increases slightly below 100~K, most likely due to the presence of paramagnetic impurities, consistent with the magnetic field dependence of magnetization at $T = 1.8$~K shown in Fig.~\ref{FIG2}(b).

Figure~\ref{FIG2}(c) shows the temperature dependence of the electrical resistivity, $\rho(T)$, of LaCd$_{3}$P$_{3}$. The $\rho(T)$ curve exhibits typical metallic behavior below 400~K, except for a distinct feature near $T_{s} = 172.5$~K. The phase transition temperature $T_{s}$ is determined from analysis of $d\rho/dT$ and indicated by the arrow in Fig.~\ref{FIG2}(d). However,  $\chi(T)$ shows no sign of a phase transition near $T_{s}$. It is notable that the resistivity at 300~K is much larger than that of typical metals, suggesting low carrier concentration in this system. The effect of a magnetic field on the phase transition is shown in Fig.~\ref{FIG2}(d), where the application of 90~kOe along the $c$ direction shifts the transition upwards by less than 1~K. It should be noted that in earlier work on polycrystalline LaCd$_{3}$P$_{3}$, $\rho(T)$ exhibited semiconducting behavior and showed no sign of a phase transition near $T_{s}$.\cite{Higuchi2016}

The temperature dependence of the specific heat, $C_{p}(T)$, of LaCd$_{3}$P$_{3}$ is shown in Fig.~\ref{FIG2}(e). $C_{p}(T)$ reveals a clear signature of the phase transition, with a $\lambda$-like anomaly at $T_{s} = 173$~K (see inset), consistent with the electrical resistivity. No thermal hysteresis is observed at $T_{s}$, as seen in the inset of Fig.~\ref{FIG2}(e). Because the specific heat curve does not follow $C_{p}(T) = \gamma T + \beta T^{3}$ at low temperatures, as shown in Fig.~\ref{FIG2}(f), neither $\gamma$ nor Debye temperature $\Theta_{D}$ can be accurately obtained. Thus, the value of $\gamma$ is estimated by a linear extrapolation to zero temperature of the $C_{p}(T)/T$ curve below 2.1~K. Within error, the estimated $\gamma$ is consistent with zero, reflecting either a small electronic enhancement or a low carrier density. Note that the $C_{p}(T)/T$ value at 1.8~K is \mbox{$\sim$ 2.5 mJ/mole K$^2$.}

\begin{figure}
\centering
\includegraphics[width=1\linewidth]{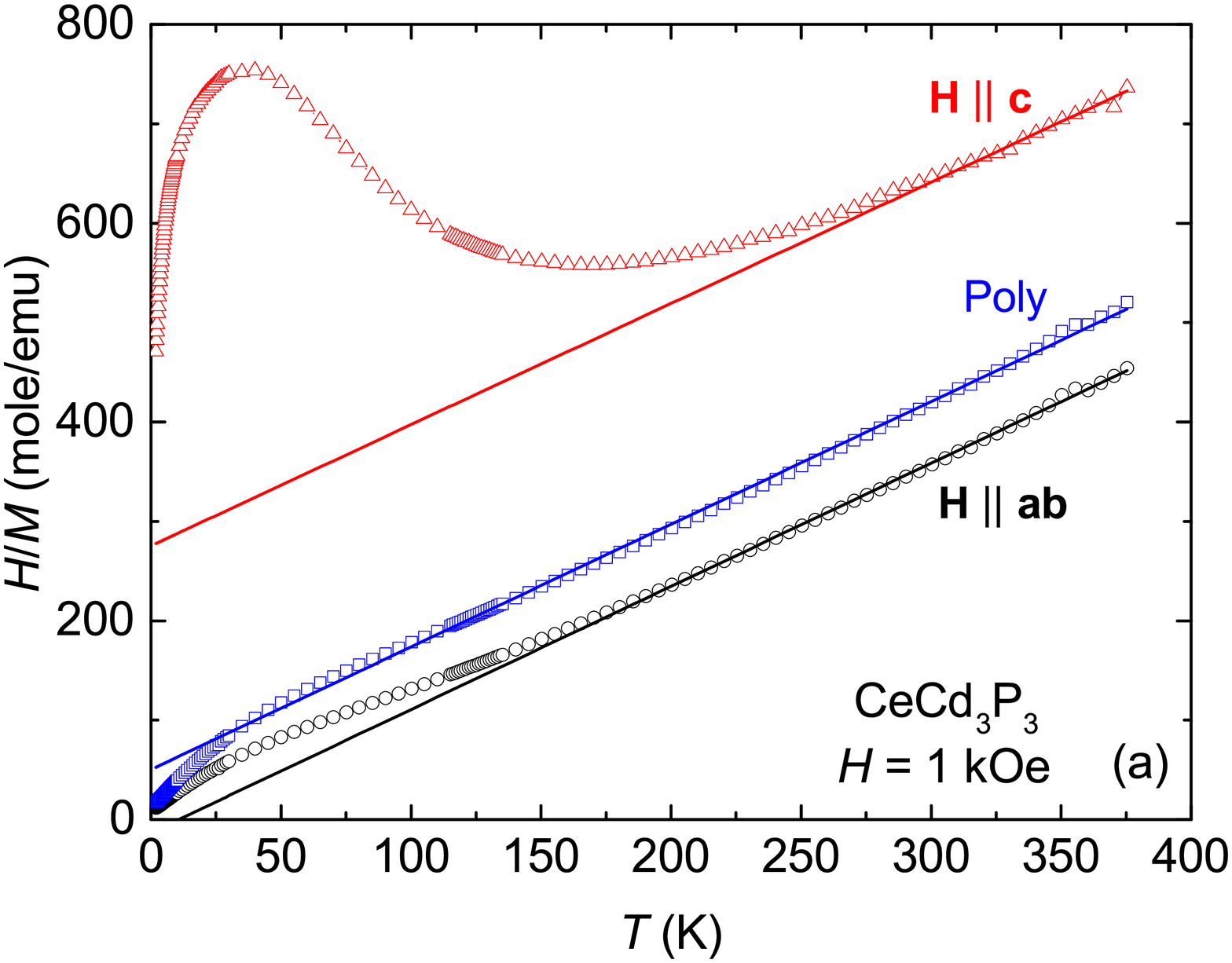}
\includegraphics[width=1\linewidth]{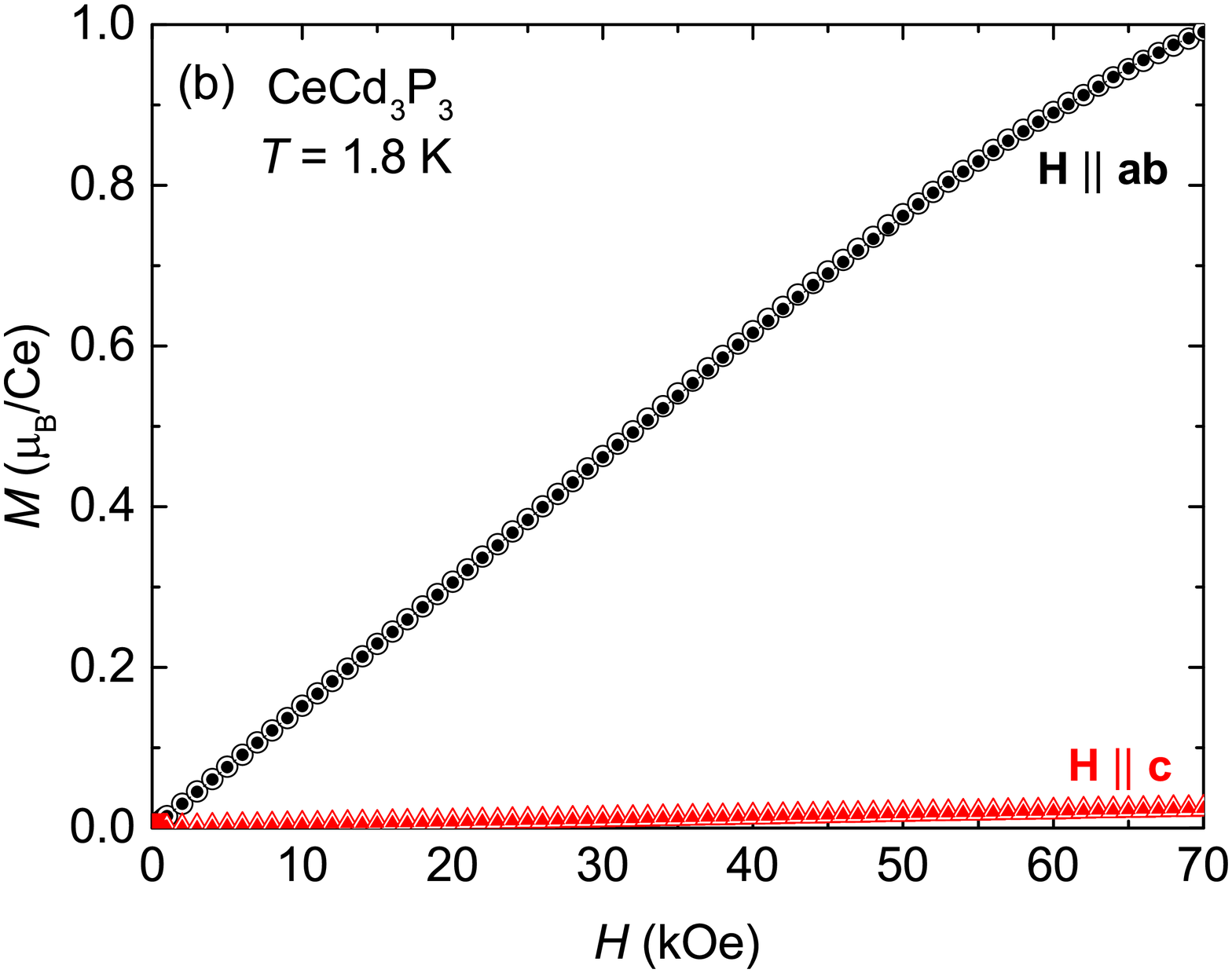}
\caption{(a) Inverse magnetic susceptibility of CeCd$_{3}$P$_{3}$ for $H {\parallel} ab$, $H {\parallel} c$, and the polycrystalline average, as defined in the text. Solid lines are   Curie--Weiss fits to the data. (b) Magnetization isotherms $M(H)$ for $H {\parallel} ab$, $H {\parallel} c$ at $T = 1.8$~K. Open and closed symbols represent up-sweeps and down-sweeps of the magnetic field, respectively.}
\label{FIG3}%
\end{figure}

The inverse magnetic susceptibility, $1/\chi(T)$, of CeCd$_{3}$P$_{3}$ is displayed in Fig.~\ref{FIG3}(a), for $H {\parallel} ab$ and $H {\parallel} c$, together with the polycrystalline average, defined by \mbox{$\chi_\mathrm{poly} = \frac{2}{3}\chi_{ab} + \frac{1}{3}\chi_{c}$}. Remarkably, $\chi_{ab}$ is much larger than $\chi_c$, reflecting two dimensional magnetic behavior most likely due to the presence of strong CEF effects. At high temperatures, the susceptibility data are well described by a Curie--Weiss law, $\chi(T) = C/(T-\theta_{p})$, where $C$ and $\theta_{p}$ are the Curie constant and Weiss temperature, respectively. The effective magnetic moments ($\mu_\mathrm{eff}$) and $\theta_{p}$ values estimated from  $1/\chi(T)$ are 2.56~$\mu_{B}$ and $-225$~K for $H {\parallel} c$, 2.54~$\mu_{B}$ and $10$~K for $H {\parallel} ab$, and 2.54~$\mu_{B}$ and $-40$~K for the polycrystalline average, respectively. Note that for $H {\parallel} c$, $\mu_\mathrm{eff}$ and $\theta_{p}$ are highly dependent on the fitting range, and the results quoted above are for a fit performed in the  range  300~K to 375~K. The effective moments obtained are close to the theoretical value of $\mu_\mathrm{eff} = 2.54$~$\mu_{B}$ for  free Ce$^{3+}$ ions. The large negative $\theta_{p}$ for $\chi_\mathrm{poly}$ indicates strong antiferromagnetic coupling in CeCd$_{3}$P$_{3}$. The deviation of magnetic susceptibility from a Curie-Weiss law below $\sim 250$~K can be attributed to CEF effects. It should be noted that $\mu_\mathrm{eff}$ and $\theta_{p}$ inferred from the polycrystalline average are consistent with earlier work on polycrystalline samples, which reported $\mu_\mathrm{eff} = 2.77~\mu_{B}$ and $\theta_{p} = -60$~K.\cite{Higuchi2016} Furthermore, these values are rather similar to findings on CeCd$_{3}$As$_{3}$\cite{Liu2016} and CeZn$_{3}$As$_{3}$\cite{Stoyko2011} powder samples. Since the magnetic susceptibility at high temperatures is strongly influenced by the CEF, $\mu_\mathrm{eff}$ and $\theta_{p}$ are also estimated by fitting the $1/\chi(T)$ curve below 10~K: $\mu_\mathrm{eff} = 2.03$~$\mu_{B}$ and $\theta_{p} = -4$~K for $H {\parallel} ab$; $\mu_\mathrm{eff} = 1.7$~$\mu_{B}$ and $\theta_{p} = -4.1$~K for the polycrystalline average. It should be noted that $1/\chi(T)$ for $H {\parallel} c$ shows no linear temperature dependence below the maximum around $\sim$ 50~K.

Figure~\ref{FIG3}(b) shows the magnetization $M(H)$ measured for $H {\parallel} c$ and $H {\parallel} ab$ in fields up to 70~kOe at $T = 1.8$~K. No hysteresis loop is observed for either orientation of magnetic field. $M(H)$ displays a large anisotropy between $H {\parallel} c$ and $H {\parallel} ab$, reflecting two-dimensional magnetic behavior, as expected from the crystal structure and the easy-($ab$) plane of magnetization. $M(H)$ for $H {\parallel} c$ is very small and  increases linearly up to 70~kOe, whereas $M(H)$ for $H {\parallel} ab$ increases linearly up to 45~kOe and starts to roll over slightly at higher magnetic fields, reaching a value of $\sim$1~$\mu_{B}$/Ce$^{3+}$ at 70~kOe. $M(H)$ at 70~kOe is smaller than that expected from the theoretical value of $gJ = 2.14~\mu_{B}$, obtained using the $J = 5/2$ free-ion result for Ce$^{3+}$.

\begin{figure}
\centering
\includegraphics[width=1\linewidth]{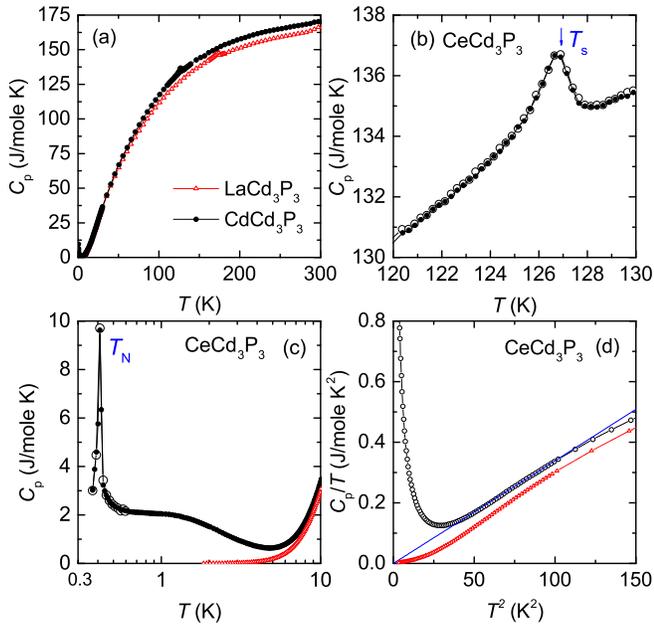}
\caption{(a) Zero field $C_{p}(T)$ for CeCd$_{3}$P$_{3}$ and LaCd$_{3}$P$_{3}$. (b) $C_{p}(T)$ of CeCd$_{3}$P$_{3}$ near the phase transition $T_{s}$. Open and closed symbols are data taken while cooling and warming, respectively. (c) $C_{p}(T)$ below 10~K on a logarithmic scale. Open and closed symbols are data taken while cooling and warming, respectively. \mbox{(d) $C_{p}/T$ vs $T^{2}$.} The solid line shows the fit to \mbox{$\gamma T + \beta T^{3}$} above 5~K. For comparison, specific heat of LaCd$_{3}$P$_{3}$ is in (c) and (d).}
\label{FIG4}%
\end{figure}

Figure~\ref{FIG4}(a) compares $C_{p}(T)$ of CeCd$_{3}$P$_{3}$ with that of LaCd$_{3}$P$_{3}$. On cooling, $C_{p}(T)$ of CeCd$_{3}$P$_{3}$ reveals a somewhat broadened $\lambda$-like feature at $T_{s} \sim 127$~K, and a sharp $\lambda$-like anomaly at $T_{N} = 0.41$~K, as seen in Figs.~\ref{FIG4}(b) and (c), respectively. No thermal hysteresis is observed at either transition. Note that previous magnetic susceptibility measurements on polycrystalline material reported no indications of magnetic ordering or electronic structure changes down to 0.48~K.\cite{Higuchi2016} In the new single crystal measurements, the anomaly at 0.41~K and the large specific heat below 5~K prevent us from using the $C_{p}(T) = \gamma T + \beta T^{3}$ analysis to estimate $\gamma$  and $\Theta_{D}$ directly from low temperature data. Instead, using a linear fit to $C_{p}/T$ vs. $T^{2}$ above 5~K, shown in Fig.~\ref{FIG4}(d), we find that the Debye temperature is $\Theta_{D} \sim 140$~K and that the electronic specific heat coefficient is consistent with zero. The negligibly small $\gamma$ value suggests either a small effective mass or low density for the charge carriers in CeCd$_{3}$P$_{3}$. 

\begin{figure}
\centering
\includegraphics[width=1\linewidth]{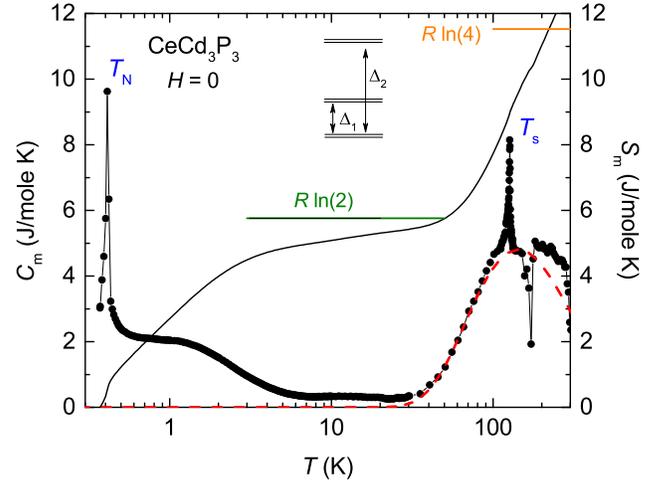}
\caption{Magnetic part of the specific heat $C_{m}$ (solid circles, left axis) and magnetic entropy $S_{m}$ (solid line, right axis). The dashed line represents the calculated Schottky contribution, based on a three-doublet configuration with $\Delta_{1} = 260$~K and \mbox{$\Delta_{2}= 600$~K}.}
\label{FIG5}%
\end{figure}

The magnetic contribution to the specific heat, $C_{m}$, of CeCd$_{3}$P$_{3}$ is obtained by subtraction of data obtained on the nonmagnetic analog LaCd$_{3}$P$_{3}$. Figure~\ref{FIG5} shows $C_{m}$ (solid circles, left axis) together with the magnetic entropy $S_{m}$ (solid line, right axis). In addition to the $\lambda$-like anomalies at $T_{N}$ and $T_{s}$, two distinct features are observed in $C_{m}$: i) a broad feature above $T_{N}$, indicative of a large electronic contribution (large $C_{m}/T$) to the magnetic specific heat; and ii) a broad local maximum at $\sim 150$~K, suggestive of a Schottky contribution. Note that the sharp features at 127~K and 173~K are due to the non-coincident structural phase transitions in CeCd$_{3}$P$_{3}$ and LaCd$_{3}$P$_{3}$. Because of the magnetic ordering below 0.41~K, an unambiguous extrapolation of the specific heat to $T = 0$ cannot be made. Thus, the integration of  $C_{m}/T$ has been performed from the base temperature of 0.37~K. This will underestimate the total magnetic entropy, especially at low temperatures. At $T_{N}$, roughly 20\% of the $R\ln(2)$ entropy is released as seen from Fig.~\ref{FIG5}. Above $T_{N}$, $S_{m}$ increases smoothly towards higher temperatures and approaches 5~J/mole-K around 5~K, which is smaller than $R\ln(2)$. When the missing entropy below 0.37~K ($\sim$1~J/mole-K) is taken into account, we believe that the true value of this entropy is $R\ln(2)$, consistent with a Kramers doublet ground state. With further increasing temperature, $S_{m}$ increases smoothly towards higher temperature, merging with the doublet-ground-state $R\ln(2)$ entropy value at about 40~K. The full $R\ln(4)$ entropy is recovered at around 200~K. The dashed line in Fig.~\ref{FIG5} represents a three-level Schottky contribution with the first excited state at 260~K and the second excited state at 600~K. Thus, the broad local maximum around 150~K can be explained by the effect of thermally excited CEF energy levels.

\begin{figure}
\centering
\includegraphics[width=1\linewidth]{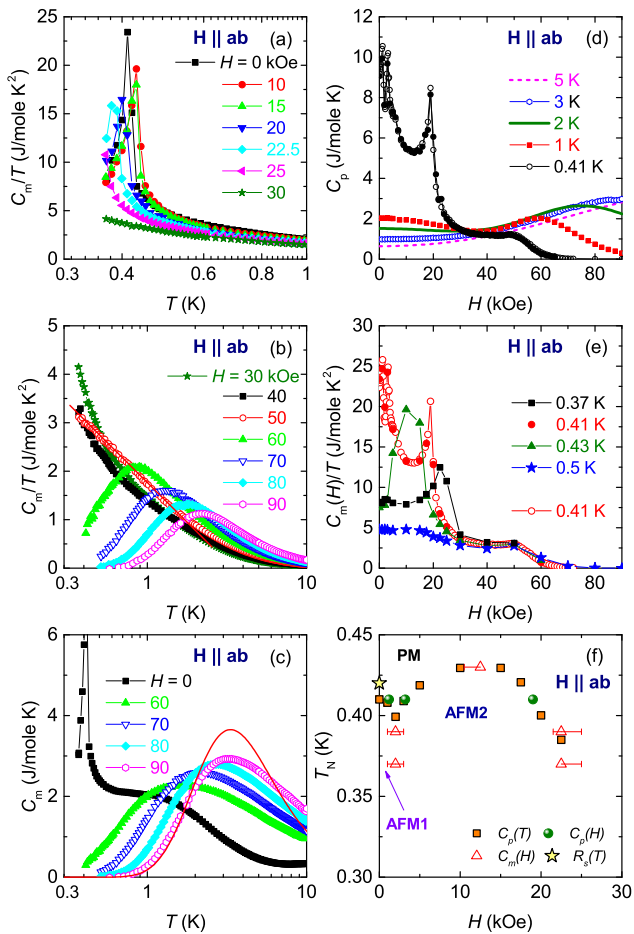}
\caption{CeCd$_{3}$P$_{3}$: Magnetic field dependence of the specific heat for $H {\parallel} ab$. (a) $C_{m}/T$ below 1~K at selected magnetic fields. (b) $C_{m}/T$ below 10~K at selected magnetic fields. Solid lines are guides to the eye. (c) $C_{m}(T)$ below 10~K at \mbox{$H$ = 0, 60, 70, 80, and 90}~kOe. The solid line represents a field-induced Schottky contribution based on two levels split by 8~K. (d) $C_{p}$ as a function of field at selected temperatures. For $T = 0.41$~K,  open and closed symbols are data taken while increasing and decreasing magnetic field. (e) $C_{m}(H)/T$ as a function of magnetic field at selected temperatures. Open circles for $T = 0.41$~K are obtained from the field dependence $C_{p}(H)$. Solid symbols are taken from the temperature dependence $C_{m}(T)$. (f) $H$--$T$ phase diagram. Solid squares and circles are taken from the peak positions in $C_{p}(T)$ and $C_{p}(H)$, respectively. Triangles are taken from the peak position in $C_{m}(H)$. Star is taken from the microwave surface resistance measurement.}
\label{FIG6}%
\end{figure}

$C_{m}/T$ curves for $H {\parallel} ab$  are plotted at select magnetic fields in Figs.~\ref{FIG6}(a) and (b). The magnetic ordering temperature $T_{N}$  increases slightly up to 15~kOe then decreases beyond this field. For $H > 22.5$~kOe, the magnetic ordering is suppressed below the base temperature of the experiment. For $H = 50$~kOe, $C_{m}/T$ increases logarithmically with decreasing temperature below  1.5~K. For $H \geq 60$~kOe, a broad maximum develops in $C_{m}/T$ and moves to higher temperature as magnetic field increases. The height and width of the maximum cannot be solely ascribed to an electronic Schottky contribution due to Zeeman splitting of the ground state doublet, as shown by the solid line in Fig.~\ref{FIG6}(c), which is calculated for an 8~K energy splitting.

The specific heat as a function of field, $C_{p}(H)$, is shown in Fig.~\ref{FIG6}(d). The $C_{p}(H)$ curve at $T = 0.41$~K indicates three peaks at $H = 1.2$, 3.2, and 19~kOe. No hysteresis is detected for these peaks. For $H > 19$~kOe, $C_{p}(H)$ drops sharply with a slope change around 50~kOe. This slope change becomes a broad local maximum and moves toward higher field as temperature increases. Figure~\ref{FIG6}(e) shows $C_{m}(H)/T$ as a function of magnetic field, extracted from the specific heat measurement as a function of temperature in a constant field. Data taken from $C_{p}(H)$ at $T = 0.41$~K (open circles) are also shown. For $H < 50$~kOe, $C_{m}(H)/T$ indicates large peaks due to magnetic ordering.  $C_{m}(H)/T$ at $T = 0.37$~K shows a maximum at $\sim$ 2~kOe and a peak at $\sim$ 22.5~kOe. When the temperature is increased to 0.43~K, a single peak is observed around $\sim 15$~kOe. At $T = 0.5$~K, $C_{m}(H)/T$  depends weakly on field below 50~kOe, but is quickly suppressed above this field. All the anomalies observed in the low temperature specific heat measurements are used to construct a partial $H$--$T$ phase diagram for $H {\parallel} ab$, shown in Fig.~\ref{FIG6}(f). Since magnetic ordering can be suppressed by external magnetic fields, it is expected that the 0.41~K phase transition in zero field is not related to ferromagnetic ordering but instead antiferromagnetism. There are at least two ordered antiferromagnetic phases, denoted AFM1 and AFM2, and a paramagnetic phase, PM.

\begin{figure}
\centering
\includegraphics[width=1\linewidth]{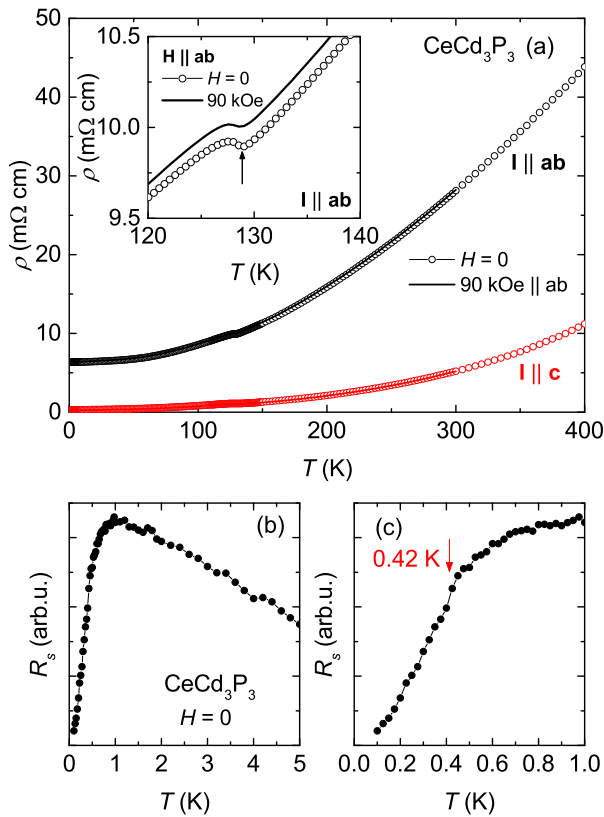}
\caption{(a) $\rho(T)$ curves for CeCd$_{3}$P$_{3}$ for currents flowing in the $ab$-plane and $c$-axis, at $H$ = 0 and 90~kOe. Inset: enlarged plot near the phase transition $T_{s}$ for $I {\parallel} ab$. Vertical arrow indicates the location of the local minimum in $\rho_{ab}(T)$ obtained, from a $d\rho/dT$ analysis. (b) Microwave surface resistance, $R_s$, below 5~K. (c) Enlarged plot of $R_s$ below 1~K. Vertical arrow indicates the phase transition temperature.}
\label{FIG7}%
\end{figure}

Figure~\ref{FIG7}(a) shows $\rho(T)$ for CeCd$_{3}$P$_{3}$, for currents flowing in the $ab$-plane ($\rho_{ab}$) and along the $c$-axis ($\rho_{c}$). The resistivity is anisotropic, with $\rho_{ab}$ about 5 times larger than $\rho_{c}$ at 300~K, where the resistivity values are 28.12~m$\Omega$-cm and  5.16~m$\Omega$-cm, respectively. $\rho(T)$ decreases with decreasing temperature, indicating metallic behavior for both current directions. There is a nonmonotonic feature around $T_{s} = 128$~K, with the local minimum in $\rho_{ab}$ determined from the zero crossing in $d\rho/dT$ and indicated by the arrow in the inset of Fig.~\ref{FIG7}(a). There is no evidence of thermal hysteresis. Since the resistivity of LaCd$_{3}$P$_{3}$ also shows a similar anomaly, around 173~K, it is likely the same phenomenon is responsible in both compounds. It should be noted that the earlier study of polycrystalline CeCd$_{3}$P$_{3}$ indicated semiconducting behavior and showed no such phase transition at $T_{s}$.\cite{Higuchi2016} A small positive magnetoresistance (MR) is observed in CeCd$_{3}$P$_{3}$ across the entire measured temperature range, and an applied magnetic field of 90~kOe does not shift $T_{s}$. Due to the large contact resistance ($\sim$50~$\Omega$ at 300~K) in the dc resistivity measurements, microwave measurements were instead performed below 5~K. The surface resistance, $R_{s}(T)$, shown in Fig.~\ref{FIG7}(b), increases with decreasing temperature and displays a slope change at 0.42~K [Fig.~\ref{FIG7}(c)]. The phase transition temperature is determined from the change in slope of $R_{s}$(T), and is consistent with the specific heat results presented earlier.

\begin{figure}
\centering
\includegraphics[width=1\linewidth]{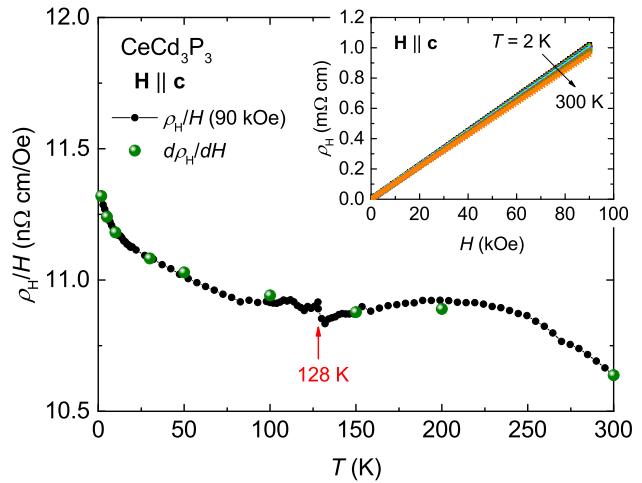}
\caption{Hall coefficient $\rho_{H}/H$ of CeCd$_{3}$P$_{3}$ at $H = 90$~kOe. Solid symbols are taken from the field dependence of Hall resistivity $\rho_{H}$. The inset shows $\rho_{H}$ curves measured at fixed temperatures of $T =$ 2, 5, 10, 30, 50, 100, 150, 200, and 300~K (top to bottom). The Hall coefficients, $d\rho_{H}/dH$, obtained from linear fits, are consistent with the $\rho_{H}/H$ temperature sweep data.}
\label{FIG8}%
\end{figure}

It is notable that $\rho(T)$ of both the La- and Ce-compounds is much larger than the resistivity values usually observed in rare-earth-based intermetallic compounds, suggesting a low carrier concentration in these systems. As a further probe  of the carrier density, the Hall resistivity $\rho_{H}$ has been measured as a function of temperature and magnetic field. $\rho_{H}$ curves for CeCd$_{3}$P$_{3}$  are plotted as a function of field  in the inset of Fig.~\ref{FIG8} at selected temperatures, where it is seen that $\rho_{H}$ is linear in field and positive for the entire temperature range. The temperature dependence of the Hall coefficient, $R_{H} = \rho_{H}/H$,  is plotted for $H = 90$~kOe in Fig.~\ref{FIG8}. It should be emphasized that $\rho_{H}/H$ is effectively temperature independent and indicates only a tiny jump at the phase transition $T_{s} = 128$~K. The positive sign of $\rho_{H}/H$ indicates that  transport is dominated by hole-like carriers. Based on a one-band model the carrier density is estimated to be $\sim 6\times10^{20}$/cm$^{3}$ at 300~K, which corresponds to $\sim$ 0.002 carriers per formula unit (f.u.), confirming the low carrier density. Thus, the negligibly small $\gamma$ values for LaCd$_{3}$P$_{3}$ and CeCd$_{3}$P$_{3}$ (obtained from the high temperature $C/T$ vs. $T^{2}$) are due to the low carrier density in these compounds.

\section{Discussion}

Due to the strong spin-orbit coupling and CEF, combined with the trigonal point symmetry of the Ce-atom, the ground state of Ce$^{3+}$ ions in CeCd$_{3}$P$_{3}$ is a Kramers doublet. Many Ce-based compounds with trigonal point symmetry show a  similar CEF scheme with a very pronounced easy-plane anisotropy and a small $c$-axis magnetization. For example, the anisotropic $\chi(T)$ curves and $\theta_{p}$ values of CeCd$_{3}$P$_{3}$ are rather similar to those of CeIr$_{3}$Ge$_{7}$\cite{Banda2018} and CeCd$_{3}$As$_{3}$\cite{Liu2016} compounds. A comprehensive analysis of the CEF scheme of Ce$^{3+}$ in the trigonal point symmetry has been presented in Ref.~\onlinecite{Banda2018}, where a strong easy-plane anisotropy is originates from large positive CEF parameters $B^{0}_{2}$ and $B^{3}_{4}$. Note that the mixing CEF parameter $B^{3}_{4}$ is absent for the sixfold point symmetry in hexagonal systems, resulting in pure $|{\pm} 1/2\rangle$, $|{\pm}3/2\rangle$, and $|{\pm} 5/2\rangle$ CEF doublets.\cite{Hutchings1964, Banda2018} Unlike the sixfold case, the presence of $B^{3}_{4}$ in trigonal symmetry induces  mixing of the $|{\pm} 5/2\rangle$ and the $|{\mp} 1/2\rangle$ states in the ground-state doublet. For isostructural CeCd$_{3}$As$_{3}$, the highly anisotropic $\chi(T)$ is well reproduced by this CEF calculation, where the energy splittings between the ground state and the first and second excited states are $\Delta_{1} = 241$~K and $\Delta_{2} = 282$~K.\cite{Banda2018} Although we have not attempted to extend the CEF calculation of Ref.~\onlinecite{Banda2018} to CeCd$_{3}$P$_{3}$, we have carried out a fit to specific heat data using the three-doublet scheme shown in Fig.~\ref{FIG5}. Based on the specific heat analysis, the CEF scheme of CeCd$_{3}$P$_{3}$ is quite similar to that of CeCd$_{3}$As$_{3}$, except for a larger overall splitting, where the second excited state is located at $\Delta_{2} = 600$~K. Since the ground state Kramers doublet is well isolated from the excited states, the low temperature thermodynamic and transport properties of CeCd$_{3}$P$_{3}$ must be governed by the low energy state of the Ce$^{3+}$ ions. Therefore, the 2D magnetism of CeCd$_{3}$P$_{3}$ cannot be explained solely on the basis of an effective $J_\mathrm{eff} = 1/2$ ground state, but both $|{\pm} 5/2\rangle$ and $|{\mp} 1/2\rangle$ contributions must be considered together.

It is notable that the high temperature anomaly at $T_{s}$ in $\rho(T)$ of $R$Cd$_{3}$P$_{3}$ is observed at the same temperature as the $\lambda$-like anomaly in the specific heat. The failure to observe a corresponding anomaly in $\chi(T)$ means that this transition cannot have a magnetic origin. In addition, the isostructural $R$Al$_{3}$C$_{3}$ ($R$ = Ce, Dy, Er, Tm, Yb, and Lu) compounds  show clear structural phase transitions.\cite{Ochiai2010} Considering that the crystal structures are of the same type as the $R$Al$_{3}$C$_{3}$ materials, it is reasonable to assume that the high temperature anomalies observed in the $R$Cd$_{3}$P$_{3}$ compounds have a structural origin. $\rho(T)$ and $\rho_{H}/H$ only show a small jump at $T_{s}$, and it is expected that the change of Fermi surface volume on passing through the phase transition will be small due to the low carrier density. Note that  $\rho(T)$ and $\rho_{H}/H$ in the low carrier density YbAl$_{3}$C$_{3}$ compound also only display a small jump at the structural phase transition.\cite{Ochiai2007} The carrier density of CeCd$_{3}$P$_{3}$ (0.002 carriers per f.u.) is about five times smaller than that of YbAl$_{3}$C$_{3}$ (0.01 carriers per f.u.).\cite{Ochiai2007} Detailed x-ray measurements of $R$Cd$_{3}$P$_{3}$ are underway to clarify the nature of the transition at $T_{s}$.

Since the specific heat contains a large electronic contribution above $T_{N}$ (the broad feature below 5~K shown in Fig.~\ref{FIG5}), an interesting question arises as to whether the 4$f$ electrons in CeCd$_{3}$P$_{3}$ are hybridized with the conduction electrons. It is a well known fact that the specific heat of many Ce- and Yb-based Kondo-lattice compounds display similar broad features at low temperatures, with large $\gamma$ values (due to the Kondo effect), accompanied by resistivities that show either maxima or logarithmic upturns resulting from Kondo scattering in conjunction with the CEF.\cite{Hewson1997} The resistivity of CeCd$_{3}$P$_{3}$ instead suggests non-hybridized metallic behavior (Fig.~\ref{FIG7}), in which the effects typically associated with a Kondo lattice system are absent. One possible explanation of the local moment behavior in CeCd$_{3}$P$_{3}$ is that there are simply not enough carriers to screen the $f$-electron moments. Supporting this interpretation, Kondo lattice compounds generally show a negative magnetoresistance at low temperatures.\cite{Hewson1997} By contrast, a small positive MR is observed in CeCd$_{3}$P$_{3}$ over the entire temperature range measured. Significantly, except for the difference in temperature of the anomalies at $T_{s}$, $\rho(T)$ in CeCd$_{3}$P$_{3}$ is the same as in LaCd$_{3}$P$_{3}$. Therefore, as there is no sign of a Kondo contribution to the resistivity, we speculate that the large electronic specific heat below 5~K is due to the effects of magnetic frustration. In a frustrated system, the presence of several competing states leads to a very large number of low-lying excitations, which manifests as an anomalously large specific heat at low temperatures.\cite{Lacroix2011} It should be noted that to rigorously exclude heavy fermion behaviour, the specific heat of CeCd$_{3}$P$_{3}$ will need to be measured down to temperatures much lower than $T_{N}$.

A further indication of the significance of frustration comes from the frustration parameter $f = |\theta_{p}/T_{N}|$, which we estimate to be of the order of 100 for CeCd$_{3}$P$_{3}$, based on the polycrystalline average $\theta_{p} \sim -40$~K and $T_{N} = 0.41$~K.  This is sufficiently large to indicate that magnetic frustration may indeed play a dominant role at low temperatures. Applying a magnetic field within the easy plane raises $T_{N}$ to higher temperatures, demonstrating partial lifting of frustration. In addition, the small amount of  magnetic entropy released at $T_{N}$ and the full $R\ln(2)$ entropy recovered at much higher temperatures indicate a competition between AFM order and frustration. Therefore, frustration effects associated with the oscillatory nature of the RKKY exchange interaction may be important in this system. It has been shown that frustrated itinerant magnets with localized $f$-moments (no Kondo effect) and a small Fermi surface display an increase of the resistivity with decreasing temperature, where the frustration is necessary to observe the resistivity upturn produced by the RKKY mechanism.\cite{Wang2016} Although due to high contact resistance in our CeCd$_{3}$P$_{3}$ samples the dc resistivity cannot be directly measured, the microwave surface resistance $R_{s}$ clearly indicates a non-logarithmic resistivity increase at low temperatures. Interestingly, a recent study of the frustrated, metallic, 2D-TL antiferromagnet PdCrO$_{2}$ found that long-range interactions such as RKKY do not compete with the spin frustration. \cite{Takatsu2009} The electrical resistivity above $T_{N}$ showed a sub-linear temperature dependence as a characteristic of the frustrated metallic magnetism, while the conduction electrons in PdCrO$_{2}$ do not strongly affect the spin frustration below $T_{N}$, which was evidenced by the 120$^{\textrm{o}}$ spin structure.

From magnetization measurements, it is clear that the spins in CeCd$_{3}$P$_{3}$ are strongly easy-plane due to the CEF, giving rise to an XY spin system. The partial $H$--$T$ phase diagram of metallic CeCd$_{3}$P$_{3}$ shown in Fig.~\ref{FIG5}(f) is similar to that of 2D insulating triangular lattice systems with easy-plane anisotropy \cite{Lee1986, Seabra2011}. A well known example of a quasi-2D easy-plane (XY) TL system is insulating RbFe(MoO$_{4}$)$_{2}$ \cite{White2013}, where the obtained magnetic phase diagram is similar to the theoretical calculation for the XY model \cite{Chubukov1991, Seabra2011, Korshunov1986, Boubcheur1996}. Based on a classical Heisenberg model for an insulating system, the phase diagram should display a magnetic structure change from a 120$^{\textrm{o}}$ structure in zero field to the up-up-down (uud) structure with increasing magnetic field, leading to a 1/3 magnetization plateau.\cite{Seabra2011} Indeed, multiple magnetic-field-induced metamagnetic transitions have been observed in many TL systems such as insulating Cs$_{2}$CuBr$_{4}$ \cite{Fortune2009} and metallic Sr$_{3}$Ru$_{2}$O$_{7}$ \cite{Ohmichi2003}. However, the interaction between spin-orbit entangled Kramers-doublet local moments on a planar triangular lattice is rather complex from a theoretical point of view.\cite{Starykh2015, Li2016, Gong2017, Zhu2018} By this analogy with insulating triangular lattice systems, it would therefore be interesting to measure magnetization below $T_{N}$ to determine whether a magnetization plateau corresponding to the uud structure exists in the CeCd$_{3}$P$_{3}$ compound. However, we suspect that the 120$^{\textrm{o}}$ magnetic order may not be stable in CeCd$_{3}$P$_{3}$. The triangular lattice will be distorted on passing through the high temperature (structural) phase transition ($T_{s}$), resulting in spatially anisotropic exchange interactions. This may resemble the case of YbAl$_{3}$C$_{3}$, where the structural phase transition from hexagonal to orthorhombic distorts the equilateral triangular lattice.\cite{Matsumura2008, Hara2012} A similar situation could occur in YbMgGaO$_{4}$, where Ga- and Mg-site mixing may destroy the 120$^{\textrm{o}}$ magnetic order and induce a quantum spin-liquid state.\cite{Parker2018, Kimchi2018}

Since the magnetic ordering in CeCd$_{3}$P$_{3}$ can be suppressed by relatively small magnetic fields, in spite of the large $\theta_{p}$, a zero temperature phase transition can be expected. The metallic nature of CeCd$_{3}$P$_{3}$ naturally introduces an interplay between RKKY and Kondo interactions. However, the hybridization between $f$-electrons and conduction electrons is very weak, which, in turn, suggests that the magnetic field will induce behavior that is distinct from that in ordinary heavy fermion systems. Taking into account the frustrated nature of the CeCd$_{3}$P$_{3}$ crystal structure and the strong AFM interactions, the ground state is expected to be degenerate. This degeneracy will be partially lifted by the high temperature (structural) phase transition $T_{s}$ and the onset of antiferromagnetic ordering below $T_{N}$. The finite specific heat up to 50~kOe at 0.5~K (Fig.~\ref{FIG6}) and unusual temperature dependence of $C_{m}/T \sim \log(1/T)$ for $H = 50$~kOe points to there being a degeneracy, implying that the frustration is not fully relieved by the phase transitions. Although CeCd$_{3}$P$_{3}$ is metallic, it is expected that the geometrically frustrated nature of the low temperature phase is the key to understanding the anomalous specific heat behavior. Recently, an experimental and theoretical effort has been underway to classify and understand the global phase diagram of AFM heavy fermion metals, where the degree of local moment quantum fluctuations can be tuned by dimensionality or geometrical frustration.\cite{Coleman2010, Si2010, Coleman2010A, Custers2012, Kim2013, Mun2013, Tokiwa2013, Fritsch2014} We expect that CeCd$_{3}$P$_{3}$, in which the Kondo coupling is negligible, will have a key role to play in developing such a phase diagram. This in turn raises questions such as: whether the long-range magnetically ordered phase in CeCd$_{3}$P$_{3}$ really displays similar physics to that of a heavy fermion system; and the nature of the interplay between magnetic frustration and the RKKY interaction. Further detailed investigations of low temperature physical properties will be necessary to address these points.

\section{Summary}

X-ray, magnetization, electrical and Hall resistivity, and specific heat measurements have been performed on single crystal $R$Cd$_{3}$P$_{3}$ ($R$ = La and Ce) compounds. The  results obtained for CeCd$_{3}$P$_{3}$ provide evidence of strongly anisotropic quasi-2D magnetism; an emergent spin-orbit entangled doublet ground state of Ce at low temperatures; a low carrier density metallic state without Kondo lattice behavior; a high temperature (structural) phase transition at $T_{s} = 127$~K; and low temperature antiferromagnetic ordering at $T_{N} = 0.41$~K. A partial \mbox{$H$--$T$} phase diagram has been constructed above 0.37~K, in which the antiferromagnetic order initially increases with magnetic field before being suppressed to lower temperatures at higher fields. The specific heat in zero field indicates a large electronic contribution ($C_{m}/T$) below $\sim 5$~K, which persists up to 50~kOe. Although it only occurs over a limited temperature range, $C_{m}/T$ at 50~kOe shows a logarithmic temperature dependence $C_{m}/T \sim \log(1/T)$. In conclusion, the complex interplay between the low carrier density metallic state  and frustrated magnetism may make CeCd$_{3}$P$_{3}$ an ideal system in which to explore strong correlation effects in a metallic host.

\section{Acknowledgements}

This work was supported by the Canada Research Chairs program, the Natural Science and Engineering Research Council of Canada, the Canadian Institute for Advanced Research, and the Canadian Foundation for Innovation.

\end{document}